\documentclass{ws-procs975x65}

\begin{document}

\title{Mass-radius relations of white dwarfs at finite temperatures}

\author{Kuantay Boshkayev,$^{1,2,*}$  Jorge A. Rueda,$^{2}$ Remo Ruffini,$^{2}$ Bakytzhan Zhami,$^{1}$ \\ Zhanerke Kalymova,$^{1}$ and Galymdin Balgimbekov.$^{1}$}

\address{$^1$Faculty of Physics and Technology, IETP, Al-Farabi Kazakh National University,\\
Al-Farabi avenue 71, Almaty, 050040, Kazakhstan\\
$^2$International Center for Relativistic Astrophysics Network,\\
Piazza della Repubblica 10, Pescara, I-65122, Italy\\
$^*$E-mail: kuantay@mail.ru}

\begin{abstract}
We construct mass-radius relations of white dwarfs taking into account the effects of rotation and finite temperatures. We compare and contrast the theoretical mass-radius relations with observational data.  
\end{abstract}


\bodymatter
\
\section{Introduction}

White dwarfs (WDs) are compact objects, formed at the final evolution stage of middle mass and low mass main sequence stars. Majority of stars in our galaxy at the end of their evolution will form WDs as an ultimate end product. Their mass is around one solar mass and size is of the same order of the Earth's size. Unlike normal stars there is no thermonuclear fusion in WDs and all the thermal energy accumulated during their formation will gradually dissipate in the form of light, heat etc. Moreover for WDs the larger the mass, the smaller the radius.\cite{shapiro1} 

In order to describe theoretically the structure and physical properties of WDs there exist, at least, three equations of state (EoS) in the literature: the well-known Chandrasekhar EoS,\cite{chandra1} the Salpeter EoS\cite{salp1,salp2} and the Relativistic Feynman-Metropolis-Teller (RFMT) EoS\cite{FMT1,rotondo,rotondo2} that generalizes both the Chandrasekhar and Salpeter EoS. All main features,  advantages, drawbacks and applications of these EoS are outlined in Ref.~\refcite{rotondo2}. Although WDs, in general, are namely investigated in classical physics, the effects of general relativity (GR) become crucial to investigate their stability close to the maximum mass (the Chandrasekhar mass limit).\cite{shapiro1,chandra1,rotondo2}

In this work we construct equilibrium configurations of static and uniformly rotating WDs using the Chandrasekhar EoS and the RFMT EoS within GR for the sake of completeness.\cite{rotondo2,bosh2013} First we perform computations for zero temperature uniformly rotating WDs at the mass shedding limit by utilizing the RFMT EoS within the Hartle formalism.\cite{H1967, HT1968, bosh2013, bosh02} Afterwards, we investigate static WDs at finite temperatures by employing the Chandrasekhar EoS. Finally, we superpose these theoretical results with the estimated data obtained from the Sloan Digital Sky Survey Data Release 4 (SDSS-E06 catalog) by Tremblay et al.\cite{tremb01}

\section{Cold Rotating and Hot Static White Dwarfs}

\begin{figure}[t]
\centerline{\includegraphics[width=0.75\columnwidth,clip]{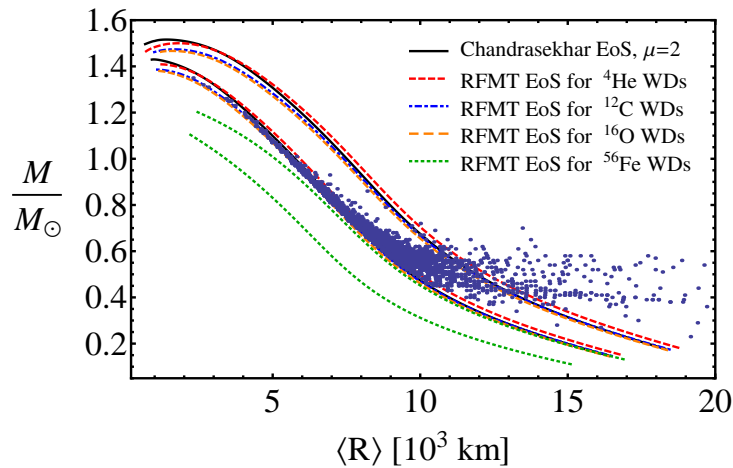}}
\caption{Mass-radius relation of uniformly rotating WDs obtained with the Chandrasekhar EoS and the RFMT EoS for T $= 0$ K case and their superposition with the estimated masses and radii of WDs taken from the SDSS-E06 catalog (blue dots, see online version). Upper curves indicate rotating WDs and lower curves indicate static WDs.}\label{fig:vs}
\end{figure}

\begin{figure}[t]
\centerline{\includegraphics[width=0.75\columnwidth,clip]{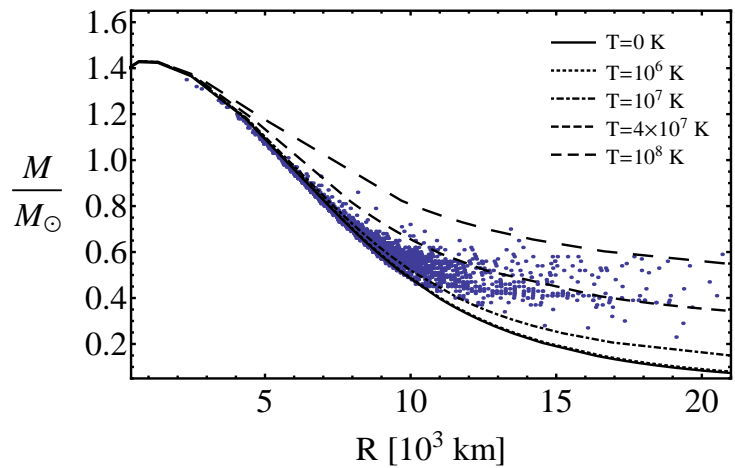}}
\caption{Mass-radius relation of static WDs obtaining using the Chandrasekhar EoS for selected finite temperatures  T = $[0,10^6,10^7,4\times10^7,10^8]$ K and their superposition with the estimated masses and radii of WDs taken from the SDSS-E06 catalog (blue dots, see online version).}\label{fig:vs2}
\end{figure}

In Fig.~\ref{fig:vs} we constructed the mass-radius relation of uniformly rotating WDs using both the Chandrasekhar and RFMT EoS in GR by fulfilling their stability criteria.\cite{rotondo2,bosh2013} The radius of the WDs in this plot is defined as the average spherical radius $\left<R\right>=(1/3) \left(R_{p}+2R_{e}\right)$, where $R_{p}$ is the polar radius, $R_{e}$ is the equatorial radius. For the sake of generality we accounted for the nuclear composition of the WD matter in the RFMT EoS. As one can see from Fig.~\ref{fig:vs} the consideration of the chemical composition along with rotation at the Keplerian limit is not sufficient to explain all the observational data. 

In order to tackle this relevant issue S. M.~de Carvalho et al.\cite{car01} proposed to include the finite-temperature effects in the EoS. Following this idea we performed a similar analysis for static WDs at finite-temperatures with the simplest Chandrasekhar EoS by solving the Tolman-Oppenheimer-Volkoff (TOV) equation (see Ref.~\refcite{bosh01} for details). The results of Ref.~\refcite{bosh01} are illustrated in Fig.~\ref{fig:vs2}. As one can see here in analogy to Ref.~\refcite{car01} the only inclusion of the finite-temperature effects in the Chandrasekhar EoS leads to the mass-radius relation in better agreement with the observational data.

It should be noted, that from the astronomical observations of isolated WDs one infers the effective surface temperature and the surface gravity but not the mass-radius relation. All main parameters are inferred and estimated by using certain models. However for WDs in close eclipsing binaries there exist techniques to measure their masses. The data obtained from these binaries are considered to be more reliable and model-independent. Therefore to perform more realistic computations one needs to take into account the effects of rotation, chemical composition and temperature together for selected WDs with known parameters. Only after one can perform more precise analyses and make further predictions.

\section{Conclusion}

In this work we calculated the mass-radius relations for cold uniformly rotating WDs within the Hartle formalism in GR using both the Chandrasekhar and RFMT EoS. We superposed our results with the estimated values of masses and radii obtained by Tremblay et al.\cite{tremb01} As a result we showed that the rotation along with the chemical composition are not sufficient enough to explain all the observational data.

Furthermore, to investigate WDs at finite-temperatures we considered a static case and solved the TOV equation numerically by using the Chandrasekhar EoS.  We compared and contrasted our results with the estimated data from the observations and showed that the data, including the range of low masses, can be covered and described by taking into account the effect of finite temperatures alone. We have considered the effects of finite-temperatures and rotation separately. For more detailed analyses one needs to consider both effects together and work with model-independent data obtained from the spectroscopic or photometric measurements of masses and radii.

In all our computations we used the uniform temperature of the WDs core. To link this temperature with a real surface temperature of a WD we exploited the Koester formula\cite{koester1} which establishes the relationship between the effective surface temperature and the core temperature. We found that most of the observed WDs core temperatures are lower than $10^8$~K  (see Fig. ~\ref{fig:vs2}). More data are still needed to confirm and extend our results. In view of recent discoveries of WDs by Koester and Kepler et al.\cite{koester2, kepler1} the theoretical study of these objects become more fascinating and overwhelming then ever before.

\section*{Acknowledgments}

This work was supported by Grant No 3101/GF4 IPC-11 and F.0679 of the Ministry of Education and Science of the Republic of Kazakhstan. B.K. acknowledges ICRANet for support and hospitality.

\end{document}